\documentclass[prl,priprint,epsf,psfig]{revtex4}
\usepackage{graphicx}
\DeclareGraphicsExtensions{.pdf,.png,.gif,.jpg}

\begin{document}

\title{Tan's universal contact and collective oscillations of strongly interacting Fermi gases}

\author{Mark DelloStritto and Theja N. De Silva\footnote{Corresponding author. Tel.: +1 607 777 3853, Fax.: +1 607 777 2546\\
E-mail address: tdesilva@binghamton.edu}}
\affiliation{Department of Physics, Applied Physics and Astronomy,
The State University of New York at Binghamton, Binghamton, New York
13902, USA.}
\begin{abstract}
We study strongly interacting two component Fermi gas near a Feshbach resonance. By using a ground state energy functional constructed based on asymptotic limits and Monte Carlo calculations, we calculate the contact, structure factor, and collective oscillation frequencies in the BCS-BEC crossover region. The calculated contact and structure factor show excellent agreements with recent experiments. We compare these results with a standard mean-field theory and find that the contact is proportional to the square of superfluid order parameter. Further, we present the chemical potential and the polytropic index in terms of homogenous energy and the contact.
\end{abstract}

\maketitle

\section{I. Introduction}

As a result of easy controllability enabled by present day laser technology, experimentalists
were able to achieve remarkable progress in trapping and cooling of atomic gases towards degenerate
temperatures~\cite{revCG}. The strongly interacting Fermi gas is one of the richest systems studied by the cold atom
community. Although harmonically trapped quantum gases are very dilute systems, most of their properties are governed by the interaction
between particles. At these characteristic densities at ultra-cold temperatures, only isotropic and
short-range \emph{s}-wave scattering between particles can take place. This scattering can be characterized by a single parameter, the \emph{s}-wave scattering length $a$. Experimentally, the scattering length can be tuned by using magnetically tuned Feshbach resonance (FB). In the unitarity limit, where $a$ is tuned to positive or negative infinity, the system is strongly interacting. At low temperatures and in the attractive interaction regime where $a < 0$, atoms form Bardeen, Cooper, and Schreifer (BCS) pairs so that the ground state is a BCS superfluid. In the repulsive interaction regime where $a > 0$, atomic potential supports two-body
molecular bound state in vacuum. In this limit the ground state corresponds to a Bose-Einstein
condensation (BEC) of these molecules. In between these two states, there is a smooth crossover
when $a$ changes its sign by passing through $\pm \infty$. At zero temperature, the only relevant dimensionless parameter for physical properties is $\xi = (k_Fa)^{-1}$, where $k_F$ is the Fermi wavevector. Because of the short range nature of the interaction between atoms in a dilute gas, the range of interatomic potential $r_0$ is much smaller than the average inter-particle spacing.

The challenging problem concerning the system properties is how to find the ground state energy $E(\xi)$. The problem does not have an analytical solution so one has to rely on numerical solutions such as Monte Carlo simulations. In general, the solutions at the limiting cases; such as $a\rightarrow 0^{\pm}$ can be obtained from the asymptotic behavior. At unitarity where $\xi = 0$, $a$ must drop out of the problem so there are only two energy scales available; Fermi energy $\epsilon_F = (3\pi^2n)^{2/3}\hbar^2/(2m)$ and the temperature $T$. Here $m$ is the mass of the atom and $n$ is the atom density in the system. Thus, dimensional analysis follows that the average single particle energy must have the form $\epsilon = \epsilon_F h(k_BT/\epsilon_F)$, where $h(x)$ is a dimensionless function. The universal constant $h(0)=3(1+\beta)/5$ has been calculated theoretically~\cite{tbeta} and is in good agreement with experiments~\cite{ebeta}. The universal many-body parameter $\beta = -0.56$ for a unitary gas and $\beta = 0$ for an ideal Fermi gas. The same argument holds for other thermodynamic quantities at unitarity. As the physical properties are independent of microscopic details, by studying a unitary Fermi gas, one learns about the
equation of state of a generic strongly interacting gas, such as high density nuclear matter found
in the center of neutron stars.

There is, however, another form of "universality" valid not only at $\xi=0$, but also in the entire BCS-BEC crossover region. This universality originated from the fact that $k_Fr_0 \ll 1$, i.e, the range of the interaction is much smaller than the inter-particle distance. Using generalized functions to deal with the singularities associated with the zero-range potential, Shina Tan recently derived a number of universal relations for Fermi gases in the BCS-BEC crossover region~\cite{tan}. These exact universal relations (Tan relations) connect microscopic properties to thermodynamic quantities. The connection is made through a single quantity termed contact ($C$), which is defined as
the high momentum ($k$) tail of the momentum distribution $n(k)$;

\begin{eqnarray}
C = \lim_{k\to \infty} k^4 n(k).
\end{eqnarray}

\noindent Remarkably, this contact carries with it all the properties and many-body physics of the system. The Tan relations are applicable for broad situations: homogeneous or trapped systems,
many-body or few-body systems, superfluid or normal phases, and finite or zero temperatures. They were later rederived by using a renormalization scheme in the quantum field theoretical framework~\cite{braaten}, by using a lattice model to regularize the singularity~\cite{werner}, by using a nonlocal quantum field theory~\cite{zhang}, and by using a Schrodinger formalism~\cite{combescot}. For a ground
state of a 4-fermion system, the contact has been calculated numerically for different values of interactions and found that the Tan relations are consistent within the systematic errors~\cite{blume}. The temperature dependence of the universal contact has been investigated by several authors using a high-temperature quantum virial expansion method~\cite{huc}, ab-initio results~\cite{palestini}, and a combined mean field-numerical method~\cite{yu}. Concurrent with the preparation of this manuscript ref.~\cite{vic} presented complementary BCS mean field results with the argument that the contact is the conjugate of the inverse scattering length. Very recently,
the Tan relations were experimentally verified at University of Colorado, Boulder, USA~\cite{jin} and at Swinburne University of Technology, Melbourne, Australia~\cite{chris}.

In this paper we investigate the strongly interacting Fermi gases at zero temperature focusing on the universal behavior. By using an energy functional $\epsilon(\xi)$ constructed by Kim \emph{et al.}~\cite{kim}, first we calculate the contact in the entire BCS-BEC crossover region. Then we calculate the structure factor from the Tan relations. Further, we find that the contact calculated from usual mean field theory is proportional to the square of the superfluid order parameter and shows very good agreement with recent experimental data and the energy functional method. In addition, we generalize the theory to trapped gases using local density approximation and polytropic equation of state. In harmonically trapped systems, we calculate the contact, energies, and collective mode frequencies. We use both hydrodynamic theory and sum rule approach and provide detail discussion about the validity of sum rules for crossover Fermi systems. All the calculated physical quantities are presented in terms of the contact and homogenous energy.

The paper is organized as follows. In the following section, we introduce the energy functional and calculate the contact for a homogenous gas as a function of $\xi$. Then we present our calculation on structure factor. In section III, we use BCS mean field theory to calculate the contact. In section IV, we generalize our theory to a harmonically trapped Fermi gas. The section V is devoted to discuss the collective oscillation frequencies and derive various oscillation frequencies. Finally, in section V, we provide a discussion and our summary.

\section{II. The energy functional and the contact}

We consider a two component Fermi gas with both spin components equally populated at zero temperature. We neglect the interaction between same spin state, but the interaction between opposite spin states is considered in the form $U(r) = (4\pi \hbar^2a/m)\delta(r)$. The total number density $n$ and the Fermi wavevector $k_F$ are related through $k_F = (3\pi^2n)^{1/3}$. The ground state energy per particle $\epsilon(\xi)$ is determined by the dimensionless interaction parameter $\xi$. As $\xi$ changes from $-\infty$ to $0$ to $+\infty$, the ground state smoothly changes from BCS superfluid state to BEC state.

In the BCS limit $a\rightarrow 0^-$, the energy per particle can be expanded in powers of $1/\xi$:

\begin{eqnarray}
\epsilon (\xi) = 2\epsilon_F \biggr(\frac{3}{10} + \frac{1}{3\pi \xi} + ....\biggr)
\end{eqnarray}

\noindent In the unitarity limit $a\rightarrow \pm \infty$, the energy per particle can be expanded in powers of $\xi$ \cite{tbeta, ebeta}:

\begin{eqnarray}
\epsilon (\xi) = \frac{3\epsilon_F (1+\beta)}{5}
\end{eqnarray}

\noindent where the universal constant $\beta$ is estimated to be $\beta = -0.56$~\cite{tbeta}. In the BEC limit $a\rightarrow 0^+$, the energy per particle can be expanded in powers of $\xi^{-3/2}$:

\begin{eqnarray}
\epsilon (\xi) = \epsilon_F \biggr[-\xi^2 + \frac{k_Fa_m}{6\pi}\biggr(1 + \frac{128 (k_Fa_m)^{3/2}}{(1350 \pi^3)^{1/2}} + .......\biggr)\biggr]
\end{eqnarray}

\noindent where $a_m \approx 0.6 a$ is the boson-boson scattering length~\cite{bs}. Here the boson is the diatomic molecule formed out of two fermions in the BEC regime. The leading term is the total binding energy density for molecules and the second term is the energy density of molecular BEC. Using the Monte Carlo results at unitarity and asymptotic results at BEC and BCS limits, the energy per particle in the entire BCS-BEC regime $\epsilon(\xi) = \epsilon_F h(\xi^{-1})$ has been constructed from a Pade parametrization by Kim \emph{et al.}~\cite{kim, kim2}. In the BCS regime where $\xi < 0$,

\begin{eqnarray}
h(x) = \frac{3}{5}-2\frac{-\delta_1 x+\delta_2 x^2}{1-\delta_3 x+\delta_4 x^2},
\end{eqnarray}

\noindent and in the BEC regime $\xi > 0$,

\begin{eqnarray}
h(x) = \frac{E_m}{2\epsilon_F}-2\frac{\alpha_1 x+\alpha_2 x^2}{1+\alpha_3 x+\alpha_4 x^2}.
\end{eqnarray}

\noindent The coefficients are found to be $\delta_1 (\alpha_1)=0.106103 (0.0316621)$, $\delta_2 (\alpha_2)=0.187515 (0.0111816)$, $\delta_3 (\alpha_3)=2.29188 (0.200149)$, and $\delta_4 (\alpha_4)=1.11616 (0.0423545)$. The molecular energy is $E_m = -\hbar^2/ma^2$. This functional has shown a very good agreement with constrained variational approximation~\cite{cva} and correctly produces the limiting behavior.

Adiabatic sweep theorem; one of the Tan's universal relations, connects the contact per unit volume (contact density) $c = C/V$ and the energy of the system

\begin{eqnarray}
c = \frac{4\pi ma^2}{\hbar^2}\frac{\partial \epsilon}{\partial a}.
\end{eqnarray}

\begin{figure}
\includegraphics[width=0.75 \columnwidth]{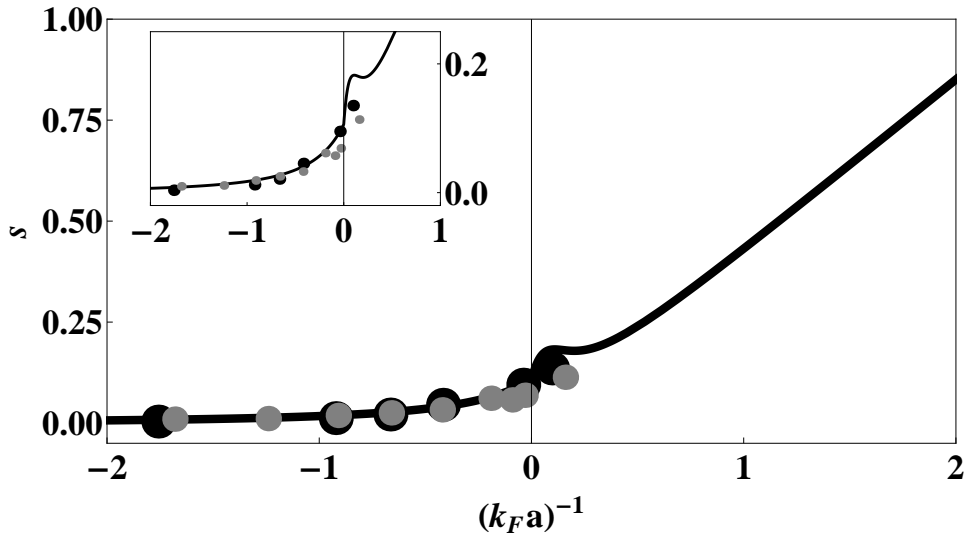}
\caption{The dimensionless contact density $s = c/k^4_F$ as a function of dimensionless interaction strength. The filled circles are experimental
data from reference~\cite{jin}. Gray filled circles: measurements from momentum distribution. Black circles: radio frequency measurements.} \label{f1}
\end{figure}

\noindent The adiabatic sweep theorem gives connection between the change in total energy density due to adiabatic change in the scattering length and the contact density. As the contact measures the probability of two antiparallel spins being close together, the adiabatic theorem links the short-range behavior to thermodynamics~\cite{braaten}. By using the energy functional $\epsilon (\xi)$, we calculate the contact density in the BCS-BEC region as a function of $\xi$. The calculated dimensionless contact density $s = c/k_F^4$ is shown in FIG~\ref{f1}. Finite temperature experimental data from ref.~\cite{jin} are shown in filled circles. While the black circles represent radio frequency measurements, the gray circles represent the momentum distribution measurements. Notice that the data in figure 2 of ref.~\cite{jin} shows a different scaling because of their data represents the contact per particle $c_p = 3 \pi^2 s$. As can be seen, our zero temperature model gives an excellent agreement with finite temperature experimental data on the BCS side of the resonance. The experiment is done at very low temperature $T = 0.11 T_F$, where $T_F$ is the Fermi temperature. Our calculation indicates that the low temperature contact is not very sensitive to the temperature. The tiny bump at the BEC side of the unitarity in our calculation is an artifact originated from the Pade approximation used to construct the energy functional. Alternatively, one can use the energy functional proposed by Manini \emph{et al}.~\cite{manini} to calculate the contact. Even though, Manini \emph{et al}.'s energy functional is more accurate on the BEC side around $\xi= 0$, it is discontinuous at $\xi =0$. Therefore, the contact shows a large jump at unitarity. In section III below, we compare the contact calculated in the present section with the BCS mean field theory. Later in section IV, we calculate the contact of harmonically trapped gas.

\subsection{Structure factor}

Tan also derived an universal relation for the large momentum behavior of the spin-antiparallel static structure factor $S_{\uparrow\downarrow}(q)\equiv S(q)$. This relation originated from the short-range behavior of the pair correlation function $n^{(2)}_{\uparrow\downarrow}(r) = \int d\textbf{R}\langle n_\uparrow(\textbf{R}-\textbf{r}/2)n_\downarrow(\textbf{R}+\textbf{r}/2)\rangle$. Taking the limit $r\rightarrow 0$, and Fourier transforming of $n^{(2)}_{\uparrow\downarrow}(r)$ gives Tan's static structure factor

\begin{eqnarray}
S(q) \simeq \frac{3\pi^2 s}{4}\frac{k_F}{q}\biggr(1-\frac{4\xi}{\pi}\frac{k_F}{q}\biggr).
\end{eqnarray}

\begin{figure}
\includegraphics[width=0.75 \columnwidth]{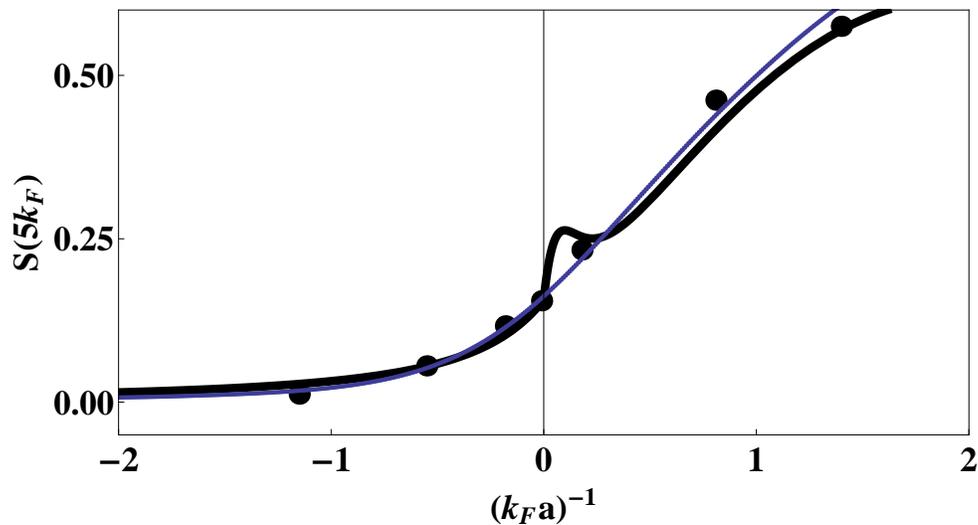}
\caption{The static structure factor $S(q)$ as a function of dimensionless interaction strength. The filled circles are experimental data from references~\cite{chris, peng}. The thin blue line is the theoretical calculation based on random phase approximation from references~\cite{hu, peng}.} \label{f2}
\end{figure}

\noindent This is the large $q \gg k_F$ behavior, however, $q$ must be smaller than the inverse of range of the interaction potential. In FIG.~\ref{f2}, we present the results for $q = 5k_F$ in the BCS-BEC crossover region. The filled circles are the experimental data from Ref.~\cite{chris} and~\cite{peng}. The thin line is the theoretical calculation based on a random phase approximation (RPA) from reference~\cite{hu} and \cite{peng}. Except at the BEC side of the unitarity, our model captures the universal power-law ($1/q$) tail of the static structure factor and shows an excellent agreement with experiments~\cite{chris} and recent theory based on RPA method~\cite{hu}.

\section{III. The contact from mean field theory}

We restrict ourselves to the wide Feshbach resonance, so that we do not need to explicitly consider the closed
channel molecule contribution. The fermions of different hyperfine states
$\uparrow$ and $\downarrow$ interact through a short-range effective
potential $U(r) = g\delta(\vec{r}\prime-\vec{r})$. The system of $N$ atoms is then described by the
Hamiltonian $H=H_1+H_2$, where $H_1$ and $H_2$ contain kinetic and interaction part of the Hamiltonian defined as,

\begin{eqnarray}
H_1 &=&
\sum_{\sigma}\int d^3\vec{r}\,
\psi_{\sigma}^{\dagger}(\vec{r})[-\frac{\hbar^{2}\nabla^2}{2m}-\mu]
\psi_{\sigma}(\vec{r}) \\ \nonumber
H_2 &= & -g\int
d^3\vec{r}\psi^{\dagger}_{\uparrow}(\vec{r})\psi^{\dagger}_{\downarrow}(\vec{r})\psi_{\downarrow}(\vec{r})
\psi_{\uparrow}(\vec{r}).
\end{eqnarray}

\noindent The field
operators $\psi_{\sigma}(\vec{r})$ obey the usual fermionic
anticommutation rules, and describe the annihilation of a fermion at
position $\vec{r}$ in the hyperfine state $\sigma$. The parameters $m$ and $\mu$ are the mass and the chemical potential of the
atoms. Using the usual BCS mean-field decoupling~\cite{bcs}, the zero temperature grand potential is given by

\begin{eqnarray}
\Omega = \sum_k\biggr[(\epsilon_k - \mu)-\sqrt{(\epsilon_k - \mu)^2+\Delta^2}\biggr] \\ \nonumber - \frac{1}{2}\Delta^2\biggr(\frac{m V}{2\pi\hbar^2a}-\sum_k\frac{1}{\epsilon_k}\biggr)
\end{eqnarray}

\noindent The BCS-Bogoliubov excitation spectrum of the atoms is given by $E_{k}=\sqrt{(\epsilon_k-\mu)^2+\Delta^2}$, where
$\epsilon_k=\hbar^2k^2/2m$ is the kinetic energy and
$\Delta=g\langle \psi_{\downarrow}(\vec{r})\psi_{\uparrow}(\vec{r})\rangle$ is the superfluid order parameter. Notice that we have already eliminated the ultraviolet divergences originated from the nature of short-range potential. This was done by regularizing the interaction term according to the relation

\begin{eqnarray}
\frac{m}{4\pi\hbar^2 a} = \frac{1}{g} + \sum_k \frac{1}{2\epsilon_k}.
\end{eqnarray}

\noindent The gap equation is obtained by the minimization of the grand potential density with respect to the superfluid order parameter,

\begin{eqnarray}
-\frac{m}{2\pi\hbar^2 a} = \frac{1}{V}\sum_k\biggr(\frac{1}{\sqrt{(\epsilon_k - \mu)^2+\Delta^2}}-\frac{1}{\epsilon_k}\biggr).
\end{eqnarray}

\noindent The fermions number density $n = N/V$ is obtained by the variation of grand potential density with respect to the chemical potential,
\begin{eqnarray}
n = \frac{1}{V}\sum_k\biggr(1-\frac{\epsilon_k-\mu}{\sqrt{(\epsilon_k - \mu)^2+\Delta^2}}\biggr).
\end{eqnarray}

\noindent Finally, evaluating $\partial \Delta/\partial a$ from the gap equation, we find the contact density

\begin{eqnarray}
c = \frac{4\pi ma^2}{\hbar^2}\biggr( \frac{\partial \Omega}{\partial a}\biggr)_{\mu, T} = \frac{m^2}{\hbar^4}\Delta^2.
\end{eqnarray}

\begin{figure}
\includegraphics[width=0.75 \columnwidth]{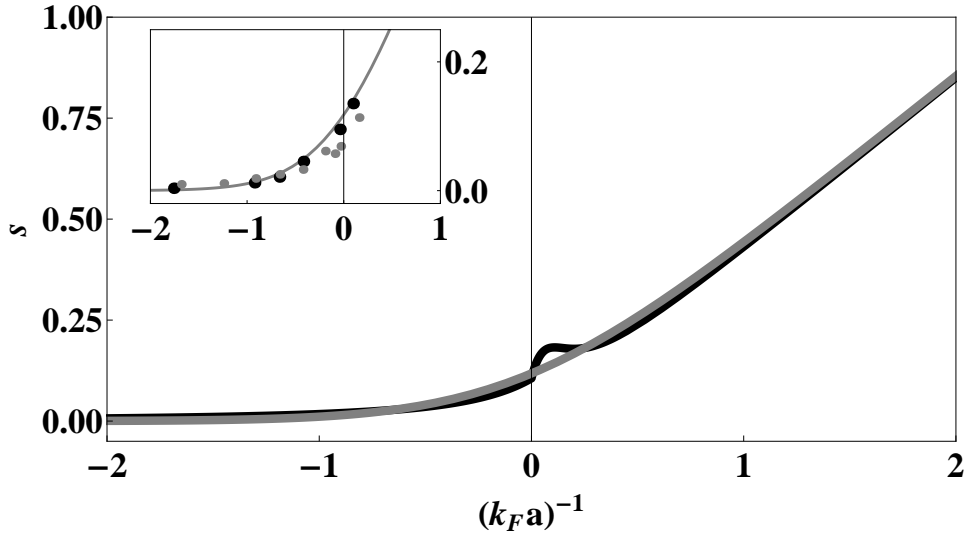}
\caption{The dimensionless contact density $s = c/k^4_F$ as a function of dimensionless interaction strength. The gray line is the BCS mean field results and black line is the results from energy functional method. The filled circles are experimental
data from reference~\cite{jin}. See FIG~\ref{f1} also.} \label{fcbcs}
\end{figure}

By converting the sum into an integral over the momentum, we numerically solve both gap equation and the number equation simultaneously to find the superfluid order parameter $\Delta$. The mean field contact density is shown in FIG.\ref{fcbcs}. As can be seen, the mean field dimensionless contact shows a reasonable agreement with our energy functional calculation carried out in the previous section. The inset shows the comparison between BCS mean field theory and the experiments. At the BCS side of the unitarity [$(k_fa)^{-1}=-0.0386$], the contact calculated from the energy functional method (0.0951) gives a remarkable agreement with the experimental value (0.0951). However, the BCS mean field result (0.1096) is off by $15\%$ at $(k_fa)^{-1}=-0.0386$. Obviously, while the contact goes to zero in the weak coupling limit ($a\rightarrow 0^-$), it approaches to infinity in the strong coupling limit ($a\rightarrow 0^+$).

\section{IV. Harmonically trapped Fermions}

Using the short-range nature of the interaction, Tan derived an additional universal relation for fermions in a harmonic trap;

\begin{eqnarray}
T -V +U = -\frac{\hbar^2}{8\pi ma}C.\label{vr1}
\end{eqnarray}

\noindent Here $T$, $V$, and $U$ are the kinetic, harmonic trapping potential, and interaction energies respectively. This generalized virial theorem was first experimentally verified at unitarity by Thomas \emph{et al}.~\cite{thomas} and later by Stewart \emph{et al}~\cite{jin} for a range of interaction strengths. In this section, we make use of generalized virial theorem in local density approximation to calculate the contact density in a trapped system.

\subsection{Chemical potential and polytropic index}

In terms of energy functional given above $\epsilon(\xi) = \epsilon_F h$, the chemical potential of the Fermi gas can be calculated by using Gibbs-Duhem relation~\cite{ma}

\begin{eqnarray}
\mu = \partial [n \epsilon(\xi)]/\partial n.
\end{eqnarray}

\begin{figure}
\includegraphics[width=0.75\columnwidth]{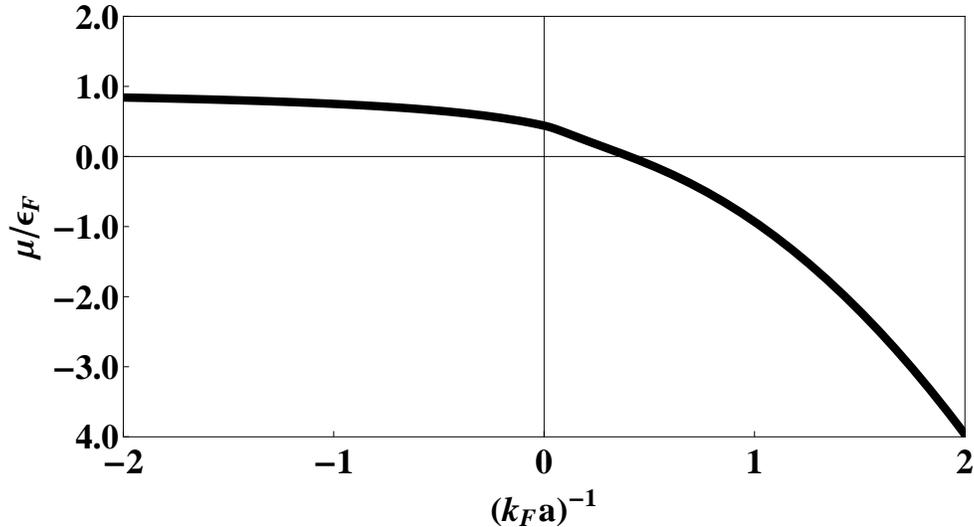}
\caption{Chemical potential $\mu$ in the BCS-BEC crossover region} \label{mu}
\end{figure}

\noindent Using this relation, we find the chemical potential $\mu = \epsilon_F (5h/3 + \pi s \xi/2)$, where $s$ is the dimensionless contact density defined above. The calculated chemical potential as a function of interaction is shown in Fig~\ref{mu}.

The polytropic index $\gamma$ relates the pressure and the density of the gas as $P \propto n^{\gamma+1}$. Alternatively, $\gamma$ relates the chemical potential as $\mu \propto n^\gamma$. We calculate this effective polytropic index as the logarithmic derivative of the chemical potential $\gamma = (n/\mu) \partial \mu/\partial n$. In the BCS-BEC crossover region, we find that the effective polytropic index reads

\begin{eqnarray}
\gamma = \frac{10h/3 + 3\pi s\xi+\pi (\partial s/\partial \xi^{-1})/2}{5h + 3\pi s\xi/2}.
\end{eqnarray}

\begin{figure}
\includegraphics[width=0.75\columnwidth]{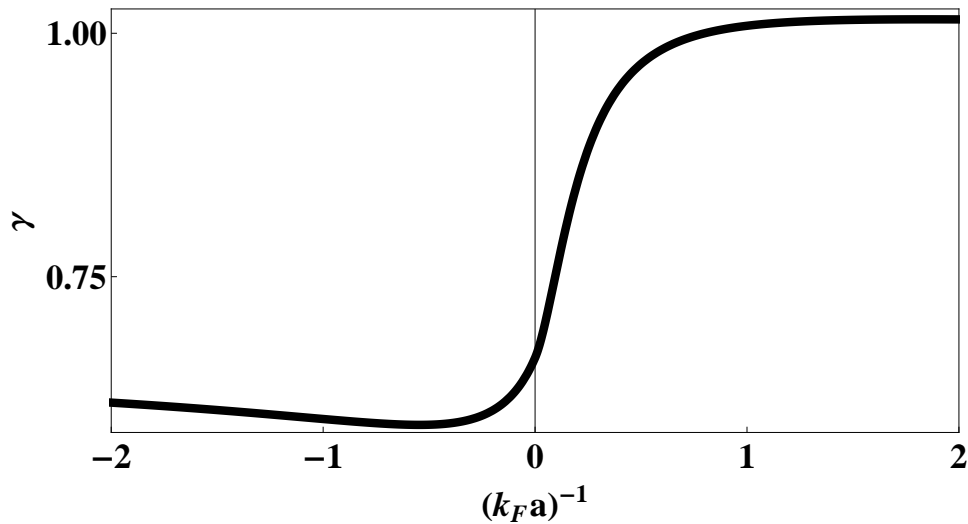}
\caption{The effective polytropic index $\gamma$ in the BCS-BEC crossover region. In the limits $(k_Fa) \rightarrow 0^{-}$ and $(k_Fa) \rightarrow \pm \infty$, the polytropic index $\gamma \rightarrow 2/3$. In the BEC limit where $(k_Fa) \rightarrow 0^{+}$, $\gamma =1$. } \label{gamma}
\end{figure}

\noindent As one expects, we find $\gamma = 2/3$ at both non interacting limit and the unitarity limit~\cite{heisel}. As we have included the molecular energy in our formalism, we find $\gamma = 0$ in the deep BEC limit. Without inclusion of the molecular contribution, we find $\gamma=1$ in the interaction dominated BEC regime. The effective polytropic index in BCS-BEC crossover region is plotted in FIG.~\ref{gamma}. In the crossover region, the effective polytropic index which excludes the molecular binding energy varies between $\gamma \sim 0.6 - 1$.

\subsection{Contact density in a trapped system}

In order to use the generalized virial theorem to find contact density in a harmonic trap, we first calculate the total trapped energy in the trap. Assuming the Fermi gas is locally homogenous, we use local density approximation and the effective polytropic index to calculate the trapped energy. Within local density approximation, the inhomogeneity is taken into account by using spatially varying chemical potential as $\mu = \mu_0-m\omega^2_0r^2/2$, where $\omega_0$ and $\mu_0$ are the trapping frequency and the chemical potential at the center of the trap. For example, the total number of atoms, $N = \int d^3r n(r)$ can be converted into integration over $\mu$ using $d\mu = -m\omega^2_0 r dr$. Assuming $\mu = A n^\gamma$, we find

\begin{eqnarray}
N = \frac{4\sqrt{2}\pi}{A^{1/\gamma}(m\omega^2)^{3/2}}\int_{-\infty}^{\mu_0}\mu^{1/\gamma}\sqrt{\mu_0-\mu}d\mu.
\end{eqnarray}

\noindent Similarly, the sum of the kinetic and interaction energies of the trapped system $T + U \equiv E = \int d^3r \epsilon[\xi(r)]$ and trapping potential energy $V = m\omega^2_0/2\int d^3r r^2 n(r)$ can be converted into the integral over $\mu$. Evaluating the integral, we find $E = (2/3\gamma) V$. Combining all the energy components, we find the total trapped energy $E_T = T + U + V$,

\begin{eqnarray}
\frac{E_T}{N} = \frac{2+3\gamma}{3\gamma}\frac{\int_{-\infty}^{\mu_0}\mu^{1/\gamma}(\mu_0-\mu)^{3/2}d\mu}{\int_{-\infty}^{\mu_0}\mu^{1/\gamma}\sqrt{\mu_0-\mu}d\mu}
\end{eqnarray}

\begin{figure}
\includegraphics[width=0.75\columnwidth]{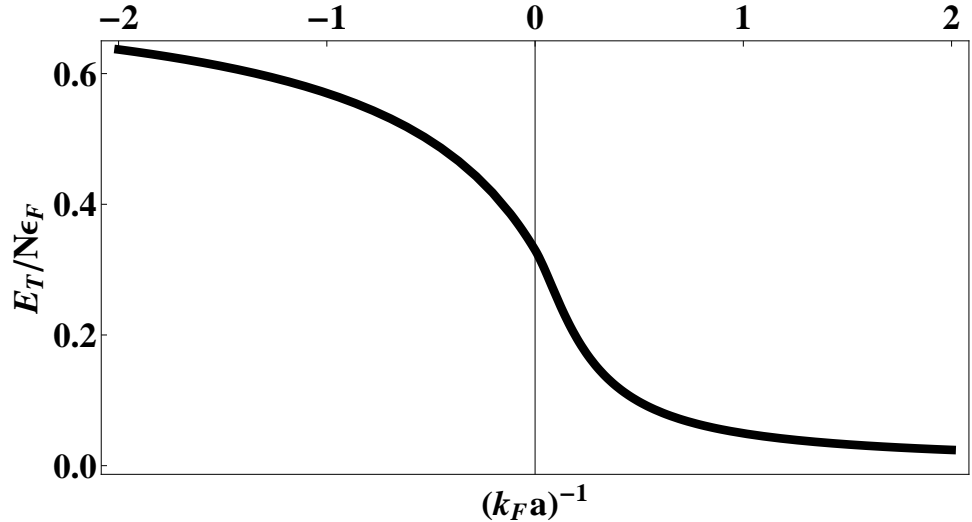}
\caption{The total energy of a harmonically trapped Fermi gas in the BCS-BEC crossover region. } \label{energy}
\end{figure}

\noindent The calculated total trapped energy in the BCS-BEC crossover region is shown in FIG.~\ref{energy}. Finally, the ratio of the contact density to the interaction strength in tapped Fermi gas is calculated from the Tan's generalized virial theorem in Eq.~(\ref{vr1}),

\begin{eqnarray}
s\xi = \frac{}{}\frac{4E_T}{3\pi N\epsilon_F}\frac{2-3\gamma}{2+3\gamma}.
\end{eqnarray}

\begin{figure}
\includegraphics[width=0.75\columnwidth]{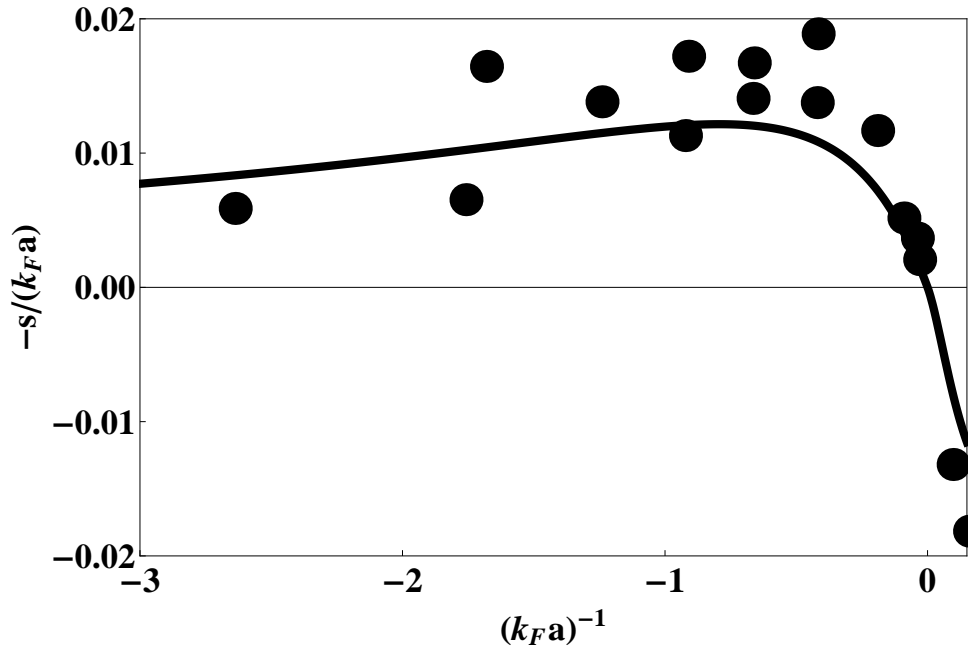}
\caption{The dimensionless contact density in a trapped Fermi gas. The filled circles are experimental
data from reference~\cite{jin}} \label{ct}
\end{figure}

\noindent The contact density of a trapped Fermi gas is shown in FIG.~\ref{ct}. The filled circles are experimental
data from reference~\cite{jin}.

\section{V. Collective oscillations}

\subsection{Hydrodynamic theory}

The dynamic of the gas can be described by the time-dependent nonlinear Schrodinger equation, known as the generalized Gross-Pitaevskii equation~\cite{pethick, ma}

\begin{eqnarray}
i\hbar \frac{\partial}{\partial t}\psi(r,t) = \biggr[\frac{\hbar^2}{2m}\nabla^2 + V(r) + \mu[n(r,t)]\biggr]\psi(r,t)\label{gp}
\end{eqnarray}

\noindent where $\psi(r,t) = \sqrt{n(r,t)} \exp[i\phi(r,t)]$ is the superfluid wave function. In terms of superfluid density $n$ and velocity $v = (\hbar/m)\nabla \phi$, Eq.~(\ref{gp}) can be converted into continuity equation and Euler equation,

\begin{equation}\label{continuity}
\frac{\partial n}{\partial t}=-\mathbf{\nabla} \cdot [n(r)
\mathbf{v}]
\end{equation}
and
\begin{eqnarray}\label{euler}
m\frac{\partial \mathbf{v}}{\partial
t}=-\mathbf{\nabla}[\frac{1}{2}m\mathbf{v}^2+V(r) + \mu(n) + qp]
\end{eqnarray}

\noindent where $qp = (-\hbar^2\nabla^2 \sqrt{n})/(2m\sqrt{n})$ is the quantum pressure term. In Thomas-Fermi approximation (TF), we assume that the gas is locally uniform so that the quantum pressure term can be neglected. This TF regime can be realized in a trap in the limit $N \rightarrow \infty$. It is interesting to note that once we drop the quantum pressure term these equations are classical so that these hydrodynamic equations are applicable for both Fermi and Bose superfluids. However, the hydrodynamic equations are sensitive to quantum corrections, statistics, and dimensionality through equation of state which enters through the density dependent local chemical potential $\mu(n)$. This hydrodynamic description is valid as long as the collisional
relaxation time is much smaller than the inverse of the collective oscillation frequencies.

For the ground state $n(r,t) = n_0(r)$ and $v(r,t) =0$. In order to study the collective oscillations above the ground state, we linearized the hydrodynamic equations by substituting $ n(r,t) = n_0(r) + \delta n(r, t)$ and  $v(r,t) = \delta v(r, t)$. After neglecting the higher order terms and keeping only the linear terms and then by combining the continuity and Euler equations, one finds the linearized version of the hydrodynamic theory

\begin{eqnarray}\label{fe1}
\frac{\partial^2 \delta n}{\partial t^2} = \mathbf{\nabla} \cdot [\frac{n_0}{m}\mathbf{\nabla} \frac{\partial \mu}{\partial n} \delta n].
\end{eqnarray}

 \noindent Taking $ \delta n(r,t) = \delta n(r) \exp[i \omega t]$ and $\mu = A n^\gamma$, this equation can be converted into a linear eigenvalue problem. In the TF regime, the continuity and Euler equations admit the ground state density $n_0(r) = n_0 (1-r^2/R_{TF}^2)^{1/\gamma}$, where the Thomas-Fermi radius $R_{TF} = \sqrt{2An_0^\gamma/(m\omega^2_0)}$. Re-scaling the length, density, and oscillation frequency by introducing $\tilde{r} = r/R_{TF}$, $\tilde{n}_0(\tilde{r}) = n_0(r)/n_0(0)$, and $\tilde{\omega} = \omega/\omega_0$ respectively, Eq.~(\ref{fe1}) can be written in dimensionless form,

\begin{eqnarray}\label{fe2}
\frac{2\tilde{\omega}^2}{\gamma}\delta n = \tilde{\mathbf{\nabla}} \cdot [\tilde{n_0} \tilde{\mathbf{\nabla}} \tilde{n}_0^{\gamma-1} \delta n].
\end{eqnarray}

\noindent For a spherically symmetric trap, the angular momentum $l$ and its projection in the z-axis $\emph{m}$ are good quantum numbers. Following the Ref.~\cite{heisel} and Ref.~\cite{yin}, the solution for the density fluctuation have the form $\delta n = R(\tilde{r}) Y_{l\emph{m}}(\theta, \phi)$, where $R(\tilde{r}) = \tilde{r}^l\tilde{n}_0^{1-\gamma}(1-\tilde{r})^{1/\gamma-1}f(\tilde{r})$. The function $f(\tilde{r})$ satisfies the hypergeometric differential equation,

\begin{eqnarray}\label{hg}
\tilde{r}(1-\tilde{r})\frac{\partial^2 f}{\partial \tilde{r}^2}+\biggr[l+3/2-(l+3/2+1/\gamma)\tilde{r}\biggr]\frac{\partial f}{\partial \tilde{r}} \\ \nonumber
+ \frac{\tilde{\omega}^2-l}{2\gamma}f.
\end{eqnarray}

\begin{figure}
\includegraphics[width=0.75\columnwidth]{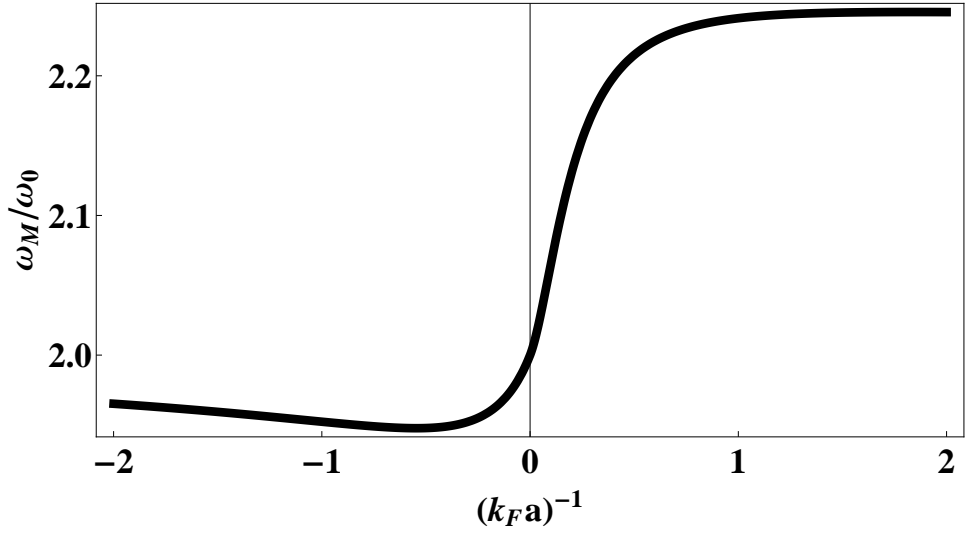}
\caption{The monopole mode frequency of a Fermi gas in a spherically symmetric trap. At weakly interacting
BCS limit $\omega_M = 2\omega_{o}$ and deep BEC limit $\omega_M = \sqrt{5}\omega_{o}$. At unitarity, mode frequencies are equal to
the ideal gas limit required by the universality.} \label{bm}
\end{figure}

\noindent The solutions of this standard equation is hypergeometric functions~\cite{heisel, yin, ma} and the eigenvalues are given by

\begin{eqnarray}\label{ev}
\tilde{\omega}_{nl}^2 = l + n_r\gamma(2n_r+2l+1) + 2n_r,
\end{eqnarray}

\noindent with radial quantum number $n_r = 0, 1, 2, ...$ and angular momentum quantum number $l = 0, 1, 2, ..$. Using the polytropic index calculated in the previous section, all the modes of collective oscillation frequencies can be presented in terms of the contact density from this energy eigenvalues. For example, monopole mode frequency or the breathing mode frequency is given when $n_r = 1$ and $l = 0$. Zero temperature monopole mode frequency $\omega_M = \omega_0\sqrt{3\gamma + 2}$ of the two component Fermi gas in a spherically symmetric trap within the
BCS-BEC crossover region is shown in FIG.~\ref{bm}. The frequencies are given as a function of the dimensionless interaction parameter $\xi$. Our model predicts $\omega_M = 2\omega_{o}$ at the weak coupling BCS limit. At the weakly repulsive BEC limit, $\omega_M = \sqrt{5}\omega_{o}$. These results are consistent with the sum rule approach present in the next subsection. At unitarity, $\omega_M = 2\omega_{o}$ as required by
universality. The other mode frequencies, such as second breathing mode ($n_r = 2$, $l = 0$) and quadrupole modes ($l =2$) can be readily calculated from Eq.~(\ref{ev}).

\subsection{Sum rule approach}

In this section, we use the sum rules approach in the linear response theory to express rigorous upper
bounds to the energies of the collective oscillations at zero temperature. The response function $\chi(z)$ at complex frequency
$z$ is written in terms of the imaginary part of the
response function $\chi^{\prime\prime}(z)$ as

\begin{eqnarray}
\chi(z) = \int\frac{d\omega}{\pi} \frac{\chi^{\prime\prime}(z)}{\omega-z}.
\end{eqnarray}

\noindent Expanding the response function in terms of $1/ z$, the moment
expansion of $\chi(z)$ is given by

\begin{eqnarray}
\chi(z) = \sum_p\frac{1}{z^p}m_{p-1}
\end{eqnarray}

\noindent where the pth moment is defined as

\begin{eqnarray}
m_p = \int\frac{d\omega}{\pi} \omega^p\chi^{\prime\prime}(\omega).
\end{eqnarray}

\noindent In general, one has to calculate the response function $\chi$ associated with the mode excitation operator to evaluate the moments. However, these moments can be computed without calculating the response function if one uses the sum rule approach~\cite{sumrule}. A major advantage of sum rule approach is that the collective mode frequencies can be calculated without the full solution of the eigenstates of the Hamiltonian. One disadvantage is that this method provides the rigorous upper bounds for the lowest collective mode frequency excited by the excitation operator. The excitation energies of the collective modes are given by the ratio $m_{p+1}/m_p$ or $\sqrt{m_{p+2}/m_p}$. When the response function is almost exhausted by a single mode (that is, the state is highly collective), then these ratio give the exact collective mode frequency. We estimate the excitation energies of the collective modes from the ratio~\cite{lc}

\begin{eqnarray}
\hbar \omega = \sqrt{\frac{m_3}{m_1}}.
\end{eqnarray}

\noindent Using the completeness relations, the moments can be written as
 \begin{eqnarray}
m_1 &=& \frac{1}{2}\langle [Q^\dagger,[H, Q]]\rangle \nonumber \\
m_3 &=& \frac{1}{2}\langle [[Q^\dagger, H],[[H,[H, Q]]]\rangle
\end{eqnarray}

\noindent where $[A, B]$ represents the commutator between the operators
A and B, $Q$ is the mode excitation operator, and $H$ is the Hamiltonian of the system

\begin{eqnarray}
H &=& \sum_i\biggr(\frac{p_i^2}{2m} + \frac{1}{2}\sum_\alpha m\omega^2_\alpha r^2_{i\alpha} \biggr) + g \sum_{i\uparrow,j\downarrow}\delta(\textbf{r}_i-\textbf{r}_j) \nonumber \\
&=& \sum_\alpha (T_\alpha + V_\alpha) + U
\end{eqnarray}

\noindent where $\alpha =x, y, z$. The excitation operators corresponding to the dipole, monopole and quadrupole modes are $Q = \sum_iz_i$,  $Q = \sum_i\textbf{r}^2_i$, and
$Q = \sum_i(\textbf{r}^2_i-3z_i^2)$ respectively. By evaluating the commutators above, the collective mode frequencies for the dipole,
monopole, and quadrupole modes are~\cite{lc}

\begin{eqnarray}
\omega_D &=& \omega_{o} \\ \nonumber
\omega_M &=& 2\omega_{o} \sqrt{\frac{T+3V+3U}{Nm\omega_0^2\langle \textbf{r}^2 \rangle}} \\ \nonumber
\omega_Q &=& 2\omega_{o} \sqrt{\frac{T+V}{4Nm\omega_0^2\langle \textbf{r}^2 \rangle}}.
\end{eqnarray}

\noindent Notice that the dipole mode is not effected by the interaction. This mode corresponds to the oscillation of the
center of mass of the atomic cloud. The calculation of monopole collective mode frequency from the first and third sum rules involves the evaluation of interaction energy $U$ of the trapped system. The calculation of this interaction energy for a crossover Fermi gas, especially at unitarity is not trivial. Indeed, the use of third sum rule for crossover fermions is problematic. For example, consider the equation of motion of the operator $Q \equiv Q_\alpha = \sum_{i}r^2_{i\alpha}$. Using Heisenberg equation of motion, one finds $\dot{Q}_\alpha = [Q_\alpha, H]/(i\hbar) = \sum_i(r_{i\alpha}p_{i\alpha} + p_{i\alpha}r_{i\alpha})/m$ and $\ddot{Q}_\alpha = 4 (T_\alpha-V_\alpha +U/2)/m$. For a spherically harmonic oscillator potential, the three terms represent the average kinetic energy $T = \langle \sum_ip_i^2/(2m)\rangle$, the harmonic oscillator potential energy $V = m\omega_{o}^2\int \textbf{r}^2n(r) d^3r/2$, and the average mean-field interaction energy $U = g\int n(r)^2d^3r/4$. In equilibrium $Q_\alpha$ is time independent, so $\ddot{Q}_\alpha = 0$. Then summing over all three components, this follows the original virial theorem,

\begin{eqnarray}
2T -2V +3U = 0.\label{vr2}
\end{eqnarray}

\noindent Now, let's combine this original virial theorem in Eq.~(\ref{vr2}) with the Tan's generalized virial theorem given in Eq.~(\ref{vr1}). For both non-interacting fermions and unitary fermions, the interaction energy $U$ turns out to be \emph{zero} at unitarity. In other words, within our sum rule approach, the third sum rule (together with original virial theorem) leads to an \emph{incorrect} interaction energy at unitarity. Regardless to this \emph{incorrect} interaction energy, the combination of these two virial theorems [Eqs. (\ref{vr1}) and (\ref{vr2})] into the sum rules produces the correct breathing mode frequency. In any case, we avoid using Eq. (\ref{vr2}), instead we use compressibility sum rule to evaluate the breathing mode frequency. Contrast to the other sum rules, the compressibility sum rule

\begin{eqnarray}
m_{-1} = \lim_{t\to 0} \int\frac{d\omega}{2\pi i\omega} e^{i\omega t}\chi^{\prime\prime}(\omega),
\end{eqnarray}

\noindent cannot be written as a set of commutators. By completing the Fourier transform over the frequency (or using the Kramers-Kronig relation), one finds that the compressibility sum rule $m_{-1} \equiv \chi(\omega = 0) = \partial Q/\partial b$ is the static response function, where $b = m\omega^2_0/2$. Using $\omega_M = \sqrt{m_1/m_{-1}}$ for the breathing mode frequency, we find

\begin{eqnarray}
\omega_M^2 = \frac{-4 \langle r^2 \rangle}{m d \langle r^2 \rangle/db}.
\end{eqnarray}

\noindent Defining $\langle r^2 \rangle = 4\pi\int_0^{r_0} r^4 n(r) dr$, where $r_0 = \sqrt{\mu_0/b}$ and using $\mu = A n^\gamma$, we find

\begin{eqnarray}
\frac{d \langle r^2 \rangle}{db} = -\frac{2 \langle r^2 \rangle}{b(3\gamma+2)}.
\end{eqnarray}

\noindent The lowest order breathing mode frequency for spherically symmetric trap is then given by $\omega_M = \omega_0\sqrt{3\gamma + 2}$. This is exactly the same results predicted by the hydrodynamic theory in previous subsection. This indicates that the rigorous upper bounds for crossover Fermi gases calculated from the sum rules are very tight. In other words, the lowest order breathing mode frequency of a trapped Fermi gas is highly collective.

\subsection{Collective modes of a Fermi gas in an axially symmetric trap}

\begin{figure}
\includegraphics[width=0.75\columnwidth]{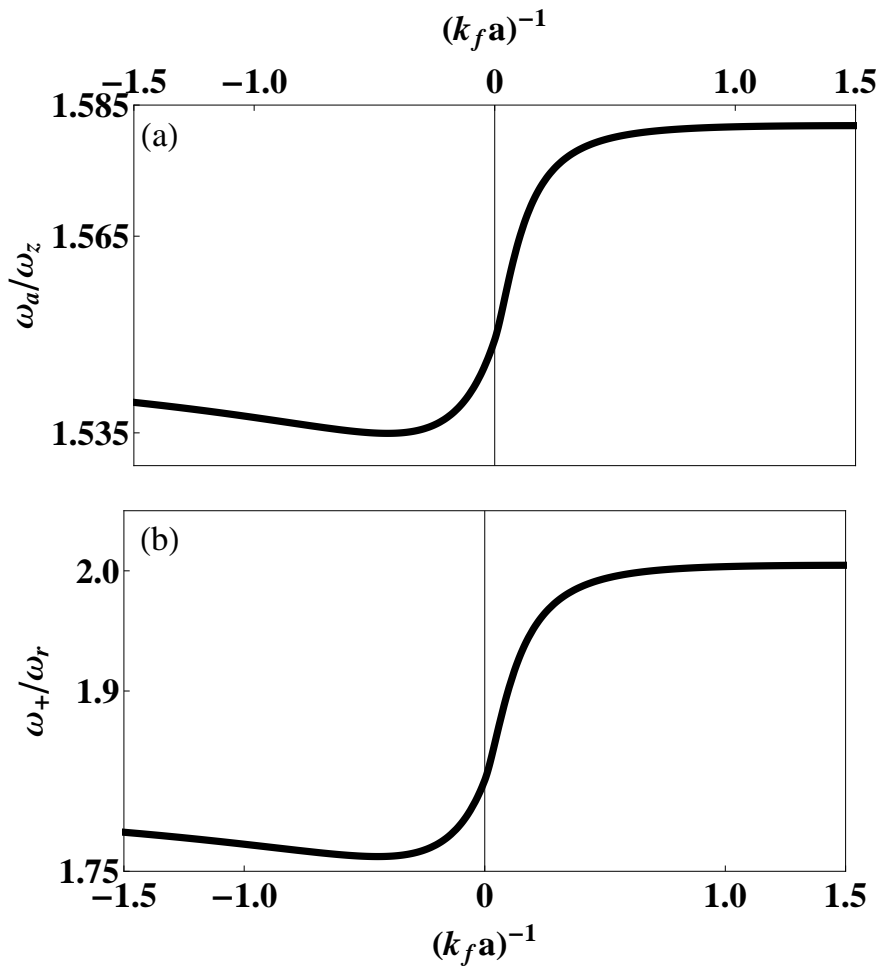}
\caption{The axial (a) and radial (b) monopole mode frequencies in the BCS-BEC crossover region.} \label{armode}
\end{figure}

In an axially symmetric trap with trapping frequency $ V(r) = m[\omega_r (x^2+y^2) + \omega_zz^2)]/2$, the monopole modes frequencies can be calculated easily using the polytropic index $\gamma$. Generalizing the hydrodynamic theory for an axially symmetric trap, the lowest frequency axial monopole mode and the radial monopole mode $\omega_a = \omega_z \sqrt{3-(\gamma+1)^{-1}}$  and $\omega_+ = \omega_r \sqrt{2(\gamma+1)}$ respectively~\cite{heisel, ma}. These expressions are valid only for a cloud with aspect ratio $\lambda \equiv \omega_z/\omega_r \ll 1$. The other limit where $\lambda \gg 1$, the monopole modes are $\omega_a = \omega_z \sqrt{2+\gamma}$  and $\omega_+ = \omega_r \sqrt{(6\gamma+4)(\gamma+2)^{-1}}$~\cite{heisel, ma}. The experimental results are available only for the case $\lambda \ll 1$~\cite{cmex1, cmex2}. These two mode frequencies in the BCS-BEC crossover region are shown in FIG.~\ref{armode}. For the limiting cases when $\gamma =1$ and $\gamma = 2/3$, the axial mode frequencies are $\omega_a = \omega_z\sqrt{5/2}$ and $\omega_a = \omega_z\sqrt{12/5}$ and the radial mode frequencies are $\omega_+ = 2\omega_r$ and $\omega_a = \omega_r\sqrt{10/3}$~\cite{lc}.

\section{VI. Summary}

By using two methods; a ground state energy functional constructed based on asymptotic limits and Monte Carlo calculations and the BCS mean field theory, we calculate the contact of a two component Fermi gas near a Feshbach resonance. The calculated contact which encapsulates all the many body physics in the BCS-BEC crossover region shows excellent agreement with the recent experiments. Within the mean field theory, we find that the contact is proportional to the square of the superfluid order parameter. We then investigate the large momentum behavior of the structure factor using a Tan's universal relation and the contact. Our structure factor calculation in the BCS-BEC crossover region also shows an excellent agreement with recent experiments and a RPA based theoretical calculation.

We combine the Tan's generalized virial theorem and local density approximation to derive the contact density in harmonically trapped fermions. Using the polytropic equation of state, we calculated the trapped energies and the collective mode frequencies of a Fermi gas trapped in a harmonic oscillator potential. We present all these quantities in terms of the homogenous energy density and the homogenous contact density.

\section{ACKNOWLEDGMENTS}

We are grateful to John Gaebler for sending us their experimental data for the contact. We thank Hui Hu for sharing with us their experimental data and RPA based theoretical results. We acknowledge Andrew Snyder for very enlightening discussions and critical comments on the manuscript.

\end{document}